\documentclass{elsart}
\usepackage{epsfig,amssymb}

\begin{document}

\begin{frontmatter}

\title{A pulsed, low-temperature beam of supersonically cooled free radical OH molecules}
\author{H.~J. Lewandowski\corauthref{cor1},}
\ead{lewandoh@jilau1.colorado.edu}
\author{Eric R. Hudson,}
\author{J.~R. Bochinski,}
\author{Jun Ye}
\corauth[cor1]{Corresponding author. Fax: 303-492-5235}

\address{JILA, National Institute of Standards and Technology and\\ University of Colorado and Department of Physics, University of Colorado,\\ Boulder, Colorado 80309-0440}

\begin{abstract}
An improved system for creating a pulsed, low-temperature
molecular beam of OH radicals has been developed. We use a pulsed
discharge to create OH from H$_2$O seeded in Xe during a supersonic
expansion, where the high-voltage pulse duration is significantly shorter
than the width of the gas pulse. The pulsed discharge allows for
control of the mean speed of the molecular packet as well as
maintains a low temperature supersonic expansion. A hot filament
is placed in the source chamber to initiate the discharge for
shorter durations and at lower voltages, resulting in a
translationally and rotationally colder packet of OH molecules.
\end{abstract}
\end{frontmatter}

% main text
\section{Introduction}
The hydroxyl radical (OH) is important in many diverse fields of
science, including ultracold collisional physics
\cite{bohn,hutson}, physical chemistry \cite{nesbitt,lester},
astrophysics \cite{astro1,astro2}, atmospheric physics
\cite{atmo}, and combustion science. Laboratory studies of OH molecules require a
controlled and a well-characterized source of the radicals.
However, because of their highly reactive nature, a controllable source is
difficult to generate in the laboratory. We present a
controllable discharge-based system to create a cold pulsed beam
of OH molecules.

There are several different techniques to produce a molecular
beam of OH radicals. The four main methods for creating OH
molecules are photolysis \cite{photo1,photo2}, radio-frequency
discharge \cite{radio}, DC discharge \cite{dc1,dc2}, and chemical
reactions \cite{chem}. We chose DC discharge because, of all of
the methods, it is the simplest and most cost-effective
technique. As presented below, the system we have developed also
fulfills the goal of producing a large sample of cold
molecules with a high phase-space density. We report several key
improvements to the standard DC discharge system, including a
pulsed high-voltage discharge to reduce heating of the molecular
packet and to allow for control of the mean speed of the molecular
packet. Also the introduction of a hot filament into the source
chamber allows the discharge to operate more stably and at a
lower voltage, thus reducing the heating
of the OH molecules during their production. Through
controlled application of a high voltage discharge pulse, we are
able to create packets of OH molecules at reasonable densities
that vary in mean speed from 265 to 470 m/s with full-width half-maximum (FWHM) velocity spread as low as 16\%.

In this Letter we describe the components of our OH source as
well as the detailed characterization of the molecular beam. Our
particular motivation for creating a cold beam of OH molecules is
to use the beam as an input to a Stark-effect based molecular
decelerator. A Stark decelerator is an apparatus that takes
advantage of the Stark energy shift, such as in the $^2\Pi_{3/2}$ (J = 3/2)
state in OH, to slow the mean
longitudinal velocity of a molecular packet to near zero 
via interactions with inhomogeneous electric
fields \cite{stark1,stark1p,stark2,stark3}.

\section{Experimental}
\begin{figure}
\leavevmode \epsfxsize=3.375in \epsffile{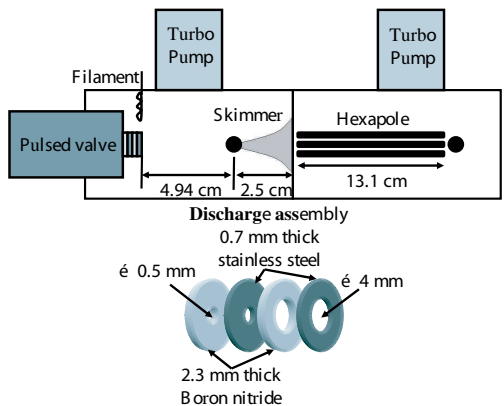}
\caption{\label{hjlfig1}Diagram of the experimental apparatus and
discharge assembly (not to scale). The system consists of two chambers
individually pumped by 300 l/s turbo pumps. A differential
pressure is maintained between the chambers by a mechanical
skimmer. A discharge assembly is mounted directly onto a pulsed
current loop actuated valve in the source chamber. The discharge assembly
consists of two disc electrodes separated by insulating spacers.
The second chamber contains an electric hexapole. Molecule
detection, by laser induced florescence, takes place in two
regions marked by black circles. }
\end{figure}

A diagram of the experimental apparatus and discharge assembly is
shown in Fig. \ref{hjlfig1}. The vacuum system consists of two
chambers separated by a mechanical skimmer, which maintains a
differential pressure between the chambers. During operation the
source (hexapole) chamber is at a pressure of 4 $\times
10^{-4}$ torr (1 $\times 10^{-6}$ torr)(1 torr = 133 Pa). A current loop actuated valve,
commercially available from R. M. Jordan Company Inc.
\cite{endore}, operates at 5 Hz to create a gas pulse $\sim$ 100
$\mu$s long. Directly in front of the 0.5 mm diameter valve nozzle is a set of stainless steel disc electrodes,
electrically isolated from one another as well as from the valve body
by Boron nitride spacers. The relevant dimensions are shown in
Fig. \ref{hjlfig1}. The electrode closest to the valve has a 0.5
mm diameter hole to match the valve nozzle. The downstream
electrode has an inner diameter of 4 mm to allow the gas to
expand as it travels between the electrodes. The valve
nozzle is placed $\sim$ 8 cm away from the downstream wall of the
vacuum chamber to ensure carrier gas atoms scattered from the
wall do not interfere with the supersonic expansion and beam
propagation.

In the second chamber, a 13 cm long electric focusing hexapole is
centered along the beam path. The hexapole is used as a tool to
determine the transverse velocity spread of the molecular beam.
The hexapole is formed by six, stainless steel, cylindrically
shaped rods with rounded ends. They are 3.18 mm in diameter and set at every
60$^\circ$ at a center-to-center radius of 4.6 mm. Alternate
rods are charged to equal magnitude but opposite polarity high
voltage.

The experimental procedure begins with the pulsed valve opening
for $\sim$100$\mu$s, thus creating a supersonically cooled pulse
of Xenon (Xe) carrier gas seeded with a few percent water. The
typical backing pressure of Xe is three atmospheres. Xe is used
instead of a lighter noble gas because of the resulting lower mean speed of the molecular beam, which is advantageous for our Stark-decelerator
application.  At a variable time after the valve opens, a high-voltage pulse is applied to the disc electrodes. The duration of
the high-voltage pulse can be varied from 1 to 200 $\mu$s. A discharge duration greater than 150 $\mu$s is considered to be essentially DC because the discharge duration is longer than the gas pulse. The
polarity of the voltage applied is such that electrons are
accelerated against the molecular beam propagation direction,
which results in a more stable discharge than the opposite
polarity. During the discharge operation, 6 A of DC current is passed though a tungsten filament, which is
located inside the source chamber. The products created by the
filament help to initiate a stable discharge at lower electrode
voltages and shorter discharge pulse durations, ultimately leading to
a colder molecular beam.

After the OH molecules are produced in the discharge, they are
allowed to fly to one of two detection regions, which are
illustrated by black circles in Fig. \ref{hjlfig1}.  The density
of OH molecules in the detection region is determined by laser-induced fluoresces (LIF).  The OH molecules are excited by a
frequency-doubled pulsed dye laser on the A$^2\Sigma_{1/2}
(v=1)\leftarrow$X$^2\Pi_{3/2} (v=0)$ transition at 282 nm. The
fluorescence from the A$^2\Sigma_{1/2}
(v=1)\rightarrow$X$^2\Pi_{3/2} (v=1)$ transition at 313 nm (with
a lifetime of 750 ns) is then imaged onto a gated photomultiplier tube
(PMT). An interference filter is placed in front of the PMT to
reduce the transmission of the excitation laser photons by $>$
10$^3$, while still allowing 15\% of the fluorescence photons to
pass. This spectral discrimination, along with careful spatial
filtering and imaging, greatly reduces the background signal from
scattered laser light. The signal from the PMT is averaged 300
times and integrated over a 3 $\mu$s time window on a digital
oscilloscope. The time from the discharge to the detection is
varied to obtain a time-of-flight (TOF) profile of the OH
molecular packet (see Fig.\ref{hjlfig2}a).

\section{Results}
\subsection{Pulsed discharge}
\begin{figure}
\leavevmode \epsfxsize=3.375in \epsffile{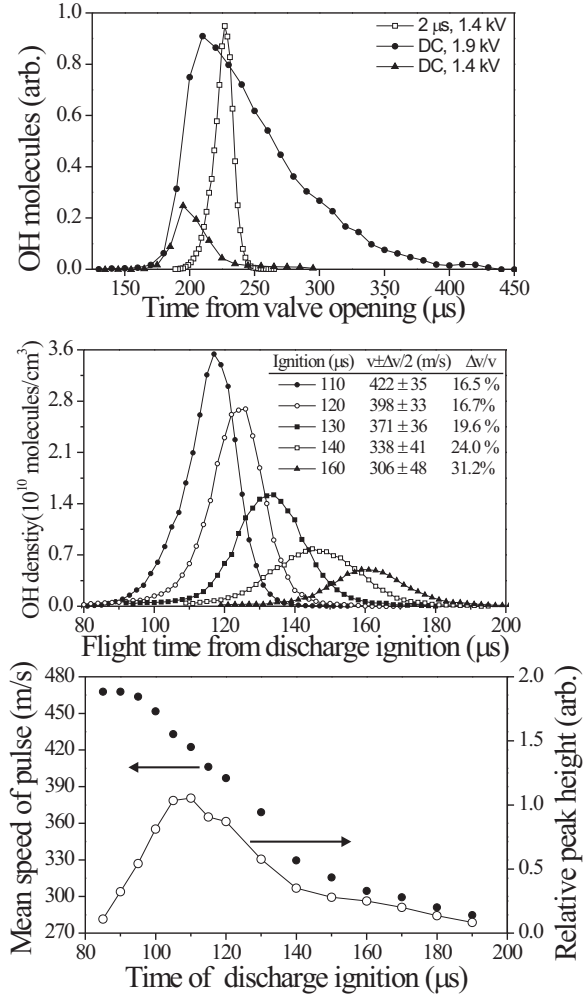}
\caption{\label{hjlfig2} (a) Longitudinal time-of-flight (TOF)
profiles, acquired in the detection region before the skimmer,
for a pulsed discharge and a DC discharge at 1.4 kV and 1.9 kV. (b) TOF
profiles, acquired in the detection region before the skimmer, as
a function of the time from the filament-assisted discharge ignition to the LIF detection. The discharge duration is 2 $\mu$s for all the data. The discharge ignition time, mean longitudinal
packet speed, and velocity width are listed in the legend for each
trace. The velocity width, $\Delta$v, is the full-width half-max
of the velocity distribution, which is determine from the TOF profiles taken at both detection locations. (c) Mean longitudinal speed
(solid circles) and relative peak height (open circles) of the
TOF profiles as a function of time from the valve trigger to the discharge ignition. For all traces, the lines serve as visual guides.}
\end{figure}

Creating OH using a high-voltage discharge pulse shorter than the
gas pulse significantly reduces the translational and the
rotational temperature, as well as permits control of the mean
speed of the OH packet. We would like to
point out that our pulsed system is distinctly different than the
one presented in \cite{dc2}, where the discharge pulse duration
is 2.5 ms, much longer than the actual gas
pulse. In our system the applied voltage between the discharging
disc electrodes is controlled by a high-voltage MOSFET switch
produced by Behlke Electronics GmbH \cite{endore}. This device can
switch up to 5 kV in well under 1 $\mu$s. Using this switch to pulse
the discharging voltage, the velocity spread of the OH
molecular packet is greatly reduced. The TOF profiles in Fig.
\ref{hjlfig2}a show a dramatic narrowing of the longitudinal
velocity distribution by reducing the duration of the discharge pulse from DC to 2 $\mu$s. Also
the measured rotational temperature of the OH beam decreases from
195 K to 28 K. The voltage on the electrodes is
increased from 1.4 kV, for the short discharge duration, to 1.9 kV for the
DC case. For a DC discharge, very few OH molecules are
produced at 1.4 kV. To make a reasonable comparison between the
two modes of operation, we increased the voltage for the DC
case until the peak signal of the OH packet was approximately
equal to that of the short discharge duration case. When the discharge is allowed to occur during the entire
gas pulse, there is a large amount of heating from the violent
discharge process. Thus shortening the discharge pulse duration greatly
reduces the temperature of the OH molecular packet and
significantly increases the molecular phase-space density.

A short discharge pulse duration also gives the freedom to produce OH
molecules at different stages during the supersonic expansion. OH  
molecular packets created at different times in the expansion
process are shown to have differing mean speeds and velocity
widths. Figure \ref{hjlfig2}b is a plot of several example TOF
profiles taken just before the skimmer where the discharge durations is 2 $\mu$s for all the data. The time between the
signal triggering the valve to open and the discharge pulse,
defined as ``ignition time,'' for each trace is listed in
the legend. There is an $\sim$ 50 $\mu$s time delay between the valve trigger and the valve opening. By timing the discharge correctly, OH molecular
packets can be created with a mean speed up to 465 m/s with the
discharge ignition at 80 $\mu$s or down to 275 m/s with the discharge ignition at
190 $\mu$s. The mean speed of the packet as a function of the discharge ignition time is summarized in Fig. \ref{hjlfig2}c .

A likely explanation for this discharge ignition-time dependent beam
velocity is that our current loop actuated valve does not operate
instantaneously or symmetrically. We see evidence of the asymmetric operation of the valve using
our pulsed discharge to sample different parts of the expanding
gas pulse. As seen in Fig. \ref{hjlfig2}c, the gas speed is
large and constant over the first 15 $\mu$s of the pulse when
the supersonic expansion has reached a steady-state beam velocity
while the number of molecules in the beam is still steadily
increasing. The speed of the gas gradually decreases as the valve
begins to close and the expansion is suboptimal. We note the peak
signal size is reached (at $\sim$ 110 $\mu$s) only after the mean
speed of the supersonic expansion beam has already decreased.
However, the FWHM longitudinal velocity spread is still a mere 16.6\%.

We confirm the asymmetric valve operation by measuring the Xe gas pulse with a fast ion gauge (FIG). The FIG trace shows a pulse with a fast leading edge and and much longer trailing edge.

For the application of a cooled molecular beam as an input to a
Stark decelerator, we require a packet of OH molecules with a high
phase-space density propagating at a low mean speed.
Choosing to create the OH molecules towards the end of the gas
pulse, for example at an ignition time of 160 $\mu$s, produces a packet
moving at an attractive mean speed of only 306 m/s. However the
amplitude of the packet is significantly smaller and the velocity
width is significantly larger than a packet created at 110 $\mu$s.
The variation of OH packet amplitudes for different ignition times can be seen in Fig. \ref{hjlfig2}c . The optimum discharge ignition time for our application is around 110 $\mu$s. For
different applications (\textit{e.g.} reactive collision dynamics), the tunability of the mean speed of the
molecular packet could be advantageous.

\subsection{Filament assisted discharge}
The other important component in the improved discharge-based
system is a hot filament in the source chamber. The hot filament
has two major effects on the discharge. First, it allows the
discharge to occur reliably and reproducibly even at the shortest
discharge pulse duration of 1 $\mu$s. The improvements
from a short discharge pulse duration are demonstrated in the previous
section. Second, the hot filament allows a stable discharge to
occur at lower voltages on the disc electrodes. Without the hot
filament, the discharge is either not stable or does not even occur at an electrode voltage
less than 3 kV; using the hot filament, the discharge is stable
down to 0.7 kV, which results in a significantly colder molecular
packet.

\begin{figure}
\leavevmode \epsfxsize=3.375in \epsffile{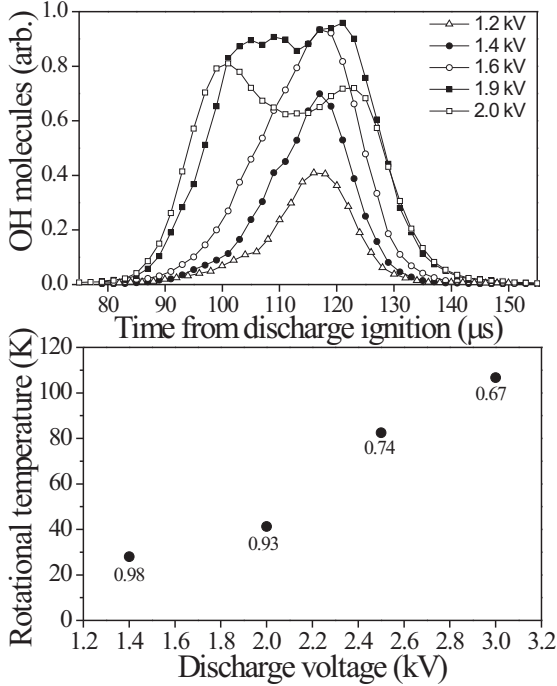}
\caption{\label{hjlfig3}(a) TOF profiles, acquired in the
detection region before the skimmer, for several discharge
voltages. The discharge pulse duration is 2 $\mu$s. The solid lines are visual guides. (b) Rotational
temperatures for different discharge voltages. The number
associated with each point is the fraction of the molecules in
the lowest rotational state (J = 3/2). }
\end{figure}

The longitudinal TOF profile and rotational temperature of the OH
molecular packet are measured for different discharge
voltages (Fig. \ref{hjlfig3}). For discharge voltages below 1.9 kV,
a single peak is observed in the TOF profile. However, for
voltages at or above 1.9 kV, the TOF profile starts to develop two
distinct maxima and indicates a considerably larger velocity
spread. We expect this heating arises from the higher energy
electrons created by a larger potential difference between the
electrodes. As the voltage is lowered from 1.6 kV to 1.2 kV, the
velocity spread remains nearly constant, but the peak number of
molecules decreases as the electrons' energy decreases and thus
creates OH molecules less efficiently. The rotational temperature
also elucidates the heating effect from the higher discharge
voltages. The rotational temperature is determined by measuring the ratio of OH molecules produced in the J = 3/2 and 5/2 states. The introduction of the hot filament permits the
reduction of the discharge voltage from 3 kV to an optimized
voltage of 1.4 kV, leading to almost a factor of four reduction
in rotational temperature (Fig. \ref{hjlfig3}b).

\subsection{Optimized conditions}
A Stark decelerator benefits from a molecular beam that has both a
high phase-space density and a low mean longitudinal speed. The
optimum configuration of the source for this application uses a 2
$\mu$s discharge duration that is ignited 110 $\mu$s after the
valve is triggered to open. The filament-assisted discharge is
created using a potential difference between the electrodes of
1.4 kV. A molecular packet created under these conditions has a
mean velocity of 422 m/s and a longitudinal velocity spread of
16.6\%, which corresponds to a translational temperature of 5 K. This is a significantly colder translational temperature than was reported by \cite{dc1} of 26 K and \cite{dc2} of 29 K.

The transverse velocity spread is determined through the use of
the hexapole focusing effect and detailed numerical simulations.
The density of OH is measured 2 mm downstream of the hexapole for different hexapole
voltages, thus producing a focusing curve. From the comparison of the numerical simulations to the hexapole focusing
data the full-width transverse velocity spread is estimated to be 35 m/s, which
corresponds to a transverse temperature of $\sim$ 1.3 K. 

The density of OH molecules just before the skimmer tip is determined from the calibrated LIF signal. The peak density of molecules in the $\Omega = 3/2$, J
= 3/2, f-component state created under these conditions is 3.5
$\times$10$^{10}$cm$^{-3}$ measured at a distance of 5 cm from the
valve nozzle. To compare with the density quoted in 
\cite{dc1}, we assume 1/r$^2$ position dependence, where r is the
distance from the nozzle, and an equal population in e and f
parity states. Our calculated density at r = 2.3 cm in both
parity states is $\sim$ 3 $\times10^{11}$cm$^{-3}$, which is a
factor of 2 less than \cite{dc1}. This lower molecular density can be attributed to a longer flight time using Xenon versus Argon. The longer flight time allows the molecular packet to spread in both the longitudinal and transverse directions reducing the density detected at a specific location.

\section{Conclusion}
In conclusion, we have developed and characterized a controllable
discharge-based source of cooled OH free radicals. Through the use
of a pulsed discharge we can tune the mean velocity of the OH beam
from 465 m/s down to 275 m/s, with a FWHM longitudinal velocity spread
as small as 16.6\%. Also the
implementation of a hot filament in the source chamber allows a
stable discharge to occur for short discharge pulse durations and at low
discharge voltages. We have shown that decreasing the
discharge pulse duration and voltage creates a colder packet of OH
molecules.


\begin{thebibliography}{11}
\bibitem{endore}
Trade names are used here for identification purposes only and do not constitute an endorsement by the authors or their institutions.
\end{thebibliography}
\end{document}